\documentclass[epj]{svjour}

\usepackage{graphicx}
\usepackage{fancyhdr}

\def\mytitle{My title} 
\def\myauthors{My name}  
\def\mytype{My type of session}
\def\mysession{My session}

\def\mytitle{Non Linear Gauge Fixing for FeynArts} 
\def\myauthors{Thomas Gajdosik}    
\def\mytype{Contributed Talk}    
\def\mysession{Colliders - SUSY Phenomenology}

\newcommand\sfrac[2]{{\textstyle \frac{#1\!\!}{#2\!\!}}}
\newcommand\ewz{\textstyle \frac{1}{\sqrt{2}}}
\begin{document}
\title{Non Linear Gauge Fixing for FeynArts}
\author{Thomas Gajdosik\inst{1,2}
\thanks{\emph{Email:} garfield@hephy.oeaw.ac.at}%
 \and
 Jurgis Pa\v{s}ukonis\inst{1}
\thanks{\emph{Email:} jurgisp@gmail.com}%
}                     
%
%
\institute{Institute of Physics, Lithuania
\and Vilnius University, Lithuania}
%
\date{}
\abstract{
We review the non-linear gauge-fixing for the Standard Model, 
proposed by F.~Boudjema and E.~Chopin, and present our implementation 
of this non-linear gauge-fixing to the Standard Model and to the minimal 
supersymmetric Standard Model in FeynArts.
} 
\maketitle
\section{Introduction}
\label{intro}

Due to the increased accuracy in the experimental area, also 
the theoretical calculations should become more accurate, even
in extensions of the SM \cite{SPA}.
The calculation in the perturbative approach requires to 
sum over all amplitudes contributing to the measured process.
There are already tools developed, that are capable of automatically 
generating and calculating Feynman diagrams at one-loop level in the 
Standard Model (SM) and the minimal supersymmetric Standard Model (MSSM)
(like \cite{feynarts}, \cite{feynartsWWWW}, \cite{grace} and others),
but together with the development of such tools there is also a need 
for ways to check the validity of the computed results, in order 
to use them with any reliability. One powerful tool to perform 
such checks arises from the procedure of gauge-fixing the theory,
as the result has to be independent of the introduced gauge-fixing 
parameters. Naturally, the more gauge-fixing parameters we have 
available, the more stringent the test becomes. 
The non-linear gauge-fixing of the Standard Model was presented in 
\cite{nonlinear}, and later fully implemented in GRACE \cite{grace}.
In this work we want to implement the same class of 
non-linear gauge-fixing in the FeynArts/FormCalc package \cite{feynarts}. 
The advantage of this package is that it is open-source and freely 
distributable.

In Sect.~\ref{sec:2} we introduce the non-linear gauge-fixing in the SM 
and we discuss the extension to the MSSM in Sect.~\ref{sec:3}. We 
describe the implementation in Feyn\-Arts in Sect.~\ref{sec:4} and 
conclude with an outlook and acknowledgements in Sect.~\ref{sec:5}.

\section{Non-linear gauge-fixing in the SM}
\label{sec:2} 

Gauge-fixing is necessary in any gauge theory, like the SM or the MSSM.
Though all fields participate in the gauge transformations, fermions 
cannot mix with the gauge bosons or the scalar bosons. Therefore we can 
ignore the fermions and their gauge transformations in the discussion 
of gauge-fixing. 

The whole theoretical issue about gauge-fixing a spontaneously 
broken Yang-Mills gauge theory can be found in text books like 
\cite{ps}, \cite{weinberg2}, or others. The remaining problem is to 
get the conventions right. In this overview we sketch only 
the conventions we use to obtain our FeynArts models files.
We write the Higgs doublet in terms of the higgs $h$ and the Goldstone
bosons $\chi$ as 
\begin{equation}
\label{eq:H-SM}  
\phi = \frac{1}{\sqrt{2}}
\left(\begin{array}{c}
-i \phi_{1} - \phi_{2} \\
i \phi_{3} + \phi_{4}
\end{array} \right)
=
\left(\begin{array}{c}
-i \chi^{+} \\
\frac{1}{\sqrt{2}} (i \chi^{0} + h )
\end{array} \right)
\enspace .
\end{equation}
The gauge group $U(1)$ with the coupling $g^{\prime}$ and the boson 
$B_{\mu}$ and the gauge group $SU(2)$ with the coupling $g$ and the 
bosons $W^{i}_{\mu}$ are mixed to 
\begin{eqnarray}
\label{eq:Amu}  
A_{\mu} &=& s_{W} W^{3}_{\mu} + c_{W} B_{\mu}
\enspace ,
\\
\label{eq:Zmu}  
Z_{\mu} &=& c_{W} W^{3}_{\mu} - s_{W} B_{\mu}
\enspace ,
\\
\label{eq:Wmu}  
W^{\pm}_{\mu} &=& \ewz
  ( W^{1}_{\mu} \mp i W^{2}_{\mu})
\enspace ,
\end{eqnarray}
with the electric charge and Weinberg angle given by
\begin{equation}
\label{eq:e-Weinberg}  
  e = \frac{g g^{\prime}}{\sqrt{g^{2} + g^{\prime 2}}}
\qquad
  s_{W} = \frac{e}{g}
\qquad
  c_{W} = \frac{e}{g^{\prime}}
\enspace .
\end{equation}
The gauge transformations on these bosons are 
\begin{eqnarray}
\label{eq:gauge-h}  
  \delta \phi_{j} 
&=& - \varphi^{a} T^{a}_{jk} \phi_{k}
\\
\label{eq:gauge-Wmu}  
  \delta W^{a}_{\mu} 
&=& \frac{1}{g_{a}} \partial_{\mu} \varphi^{a}
  + f^{abc} W^{b}_{\mu} \varphi^{c}
\enspace ,
\end{eqnarray}
where the $U(1)$ gauge group gets the index $4$: $g_{4} = g^{\prime}$ 
and $f^{abc}$ is the structure constant of $SU(2)$ for $a,b,c < 4$ and
zero otherwise. $[ T^{a} , T^{b} ] = f^{abc} T^{c}$ are real.

Gauge fixing with the Faddeev Popov procedure gives the gauge-fixing 
part $\mathcal{L}_{\mathrm{gf}}$ and the ghost part 
$\mathcal{L}_{\mathrm{gh}}$ of the Lagrangian in terms of the 
gauge-fixing functions
\begin{eqnarray}
\label{eq:GA}  
  F^{A}  
&=& 
  \partial^{\mu} A_{\mu}
\enspace ,
\\
\label{eq:GZ}  
  F^{Z}  
&=&
  \partial^{\mu} Z_{\mu}
- \xi_{Z} m_{Z} \chi^{0} 
- \tilde{\varepsilon} \frac{e \xi_{Z}}{2 s_{W} c_{W}} (h \chi^{0})
\enspace ,
\\
\label{eq:GW}  
  F^{+}  
&=&
  \partial^{\mu} W^{+}_{\mu}
- \xi_{W} m_{W} \chi^{+} 
- \frac{e \xi_{Z}}{2 s_{W}} 
  ( \tilde{\delta} h - i \tilde{\kappa} \chi^{0}) \chi^{+} 
\nonumber \\ & &
- i e ( \tilde{\alpha} A^{\mu} 
      + \tilde{\beta} \sfrac{c_{W}}{s_{W}} Z^{\mu} ) W^{+}_{\mu}
\enspace ,
\end{eqnarray}
and $F^{-} = (F^{+})^{*}$, expressed in the physical fields as
\begin{eqnarray}
\label{eq:Lgf}
  \mathcal{L}_{\mathrm{gf}}  
&=& 
- \frac{1}{2 \xi_{A}} (F^{A})^{2}
- \frac{1}{2 \xi_{Z}} (F^{Z})^{2}
- \frac{1}{\xi_{W}} F^{+} F^{-}
\, , \enspace
\\
\label{eq:Lgh}
  \mathcal{L}_{\mathrm{gh}}
&=&
  \bar{c}^{a}
  \left(\left. 
    \frac{\delta F^{a}}{\delta \alpha^{b}} 
  \right|_{\alpha = 0} \right) 
  c^{b}
\, ,
\end{eqnarray}
where the ghost and the gauge parameters $\varphi^{a}$ have to be 
transformed into a physical basis. For the exact procedure see \cite{jp}.

Obviously there are additional vertices coming from the non-linear part
of the gauge-fixing functions. These vertices are proportional to the 
additional gauge parameters $\tilde{\alpha}$, $\tilde{\beta}$, 
$\tilde{\delta}$, $\tilde{\varepsilon}$, and $\tilde{\kappa}$. 

\section{Non-linear gauge-fixing in the MSSM}
\label{sec:3} 

In the MSSM the gauge bosons eqs.(\ref{eq:Amu}, \ref{eq:Zmu}, 
\ref{eq:Wmu}) are the same, but we have two higgs doublets instead of 
eq.(\ref{eq:H-SM}):
\begin{eqnarray}
\label{eq:H1}
H_{1} &=& 
\left(\begin{array}{c}
  \frac{1}{\sqrt{2}} 
  [ ( O_{1h} + i s_{\beta} O_{3h} ) H_{h} 
  - i c_{\beta} G^{0} ] \\
    s_{\beta} H^{-} - c_{\beta} G^{-} 
\end{array} \right)
\enspace ,
\\
\label{eq:H2}
H_{2} &=& 
\left(\begin{array}{c}
    c_{\beta} H^{+} + s_{\beta} G^{+} \\
  \frac{1}{\sqrt{2}} 
  [ ( O_{2h} + i c_{\beta} O_{3h} ) H_{h} 
  + i s_{\beta} G^{0} ] 
\end{array} \right)
\enspace ,
\end{eqnarray}
where $H_{h}$ are the three neutral higgs bosons and 
\begin{equation}
\label{eq:tanb}  
\frac{s_{\beta}}{c_{\beta}}=\frac{v_{2}}{v_{1}} = \tan \beta
\end{equation}
is the ratio of the vaccum expectation values of the higgs fields. In 
the normal MSSM without induced CP violation in the Higgs sector 
$O_{3h} = \delta_{3h}$, $H_{3} = A^{0}$, and $O_{jk}$ describes the 
normal mixing of the two neutral CP even higgs fields $h^{0}$ and 
$H^{0}$. 

The definition of the parts of the Lagrangian eq.(\ref{eq:Lgf}) and
eq.(\ref{eq:Lgh}) stay the same, but the gauge-fixing functions
eq.(\ref{eq:GZ}) and eq.(\ref{eq:GW}) are changed to account for the 
extended Higgs sector:
\begin{eqnarray}
\label{eq:GZ-MSSM}  
  F^{Z}  
&=&
  \partial^{\mu} Z_{\mu}
+ \xi_{Z} m_{Z} G^{0} 
+ \frac{e \xi_{Z}}{2 s_{W} c_{W}} (\tilde{\varepsilon}_{h} H_{h} G^{0})
\enspace ,
\\
\label{eq:GW-MSSM}  
  F^{\pm}  
&=&
  \partial^{\mu} W^{+}_{\mu}
+ i \xi_{W} m_{W} G^{+} 
+ \frac{i e \xi_{W}}{2 s_{W}} 
  ( \tilde{\kappa} G^{0} \mp i \tilde{\delta}_{h} H_{h} ) G^{+} 
\nonumber \\ & &
+ i e \xi_{W} ( \tilde{\alpha} A^{\mu} 
      + \tilde{\beta} \sfrac{c_{W}}{s_{W}} Z^{\mu} ) W^{+}_{\mu}
\enspace .
\end{eqnarray}
Writing the gauge transformations eq.(\ref{eq:gauge-h}) and 
eq.(\ref{eq:gauge-Wmu}) in terms of the physical fields 
\begin{eqnarray}
\label{eq:gauge:Amu}  
  \delta A_{\mu}  
&=&
- \partial_{\mu} \varphi^{A}
+ i e 
  ( W^{-}_{\mu} \varphi^{+}
  - W^{+}_{\mu} \varphi^{-} )
\enspace ,
\\
\label{eq:gauge:Zmu}  
  \delta Z_{\mu}  
&=&
- \partial_{\mu} \varphi^{Z}
+ i e \sfrac{c_{W}}{s_{W}}
  ( W^{-}_{\mu} \varphi^{+}
  - W^{+}_{\mu} \varphi^{-} )
\enspace ,
\\
\label{eq:gauge:Wmu}  
  \delta W^{\pm}_{\mu}  
&=&
- \partial_{\mu} \varphi^{\pm}
\mp i e ( A_{\mu} + \sfrac{c_{W}}{s_{W}} Z_{\mu} ) \varphi^{\pm}
\nonumber \\ & & \enspace
\pm i e ( \varphi^{A} + \sfrac{c_{W}}{s_{W}} \varphi^{Z} ) W^{\pm}_{\mu}
\enspace ,
\\
\label{eq:gauge:H0} 
  \delta H_{h}  
&=&
  \sfrac{e}{s_{W}} O^{cs}_{h}
  ( \sfrac{1}{2 c_{W}} G^{0} \varphi^{Z}
  + i G^{+} \varphi^{-} - i G^{-} \varphi^{+} )
\nonumber \\ & &
- \sfrac{i e}{s_{W}} 
  ( s_{\beta} O_{1h} - c_{\beta} O_{2h} ) 
  ( H^{+} \varphi^{-} - H^{-} \varphi^{+} )
\nonumber \\ & &
+ \sfrac{e}{s_{W}} O_{3h} 
  ( H^{+} \varphi^{-} + H^{-} \varphi^{+} )
\enspace ,
\\
\label{eq:gauge:G0} 
  \delta G^{0}  
&=&
- ( m_{Z} + \sfrac{e}{2 s_{W} c_{W}} O^{cs}_{h} H_{h} ) \varphi^{Z}
\nonumber \\ & &
+ \sfrac{e}{s_{W}} ( G^{+} \varphi^{-} + G^{-} \varphi^{+} )
\enspace ,
\\
\label{eq:gauge:Gpm} 
  \delta G^{\pm}  
&=&
\pm i ( m_{W} 
      + \sfrac{e}{2 s_{W}} 
        ( O^{cs}_{h} H_{h} \pm i G^{0} ) ) \varphi^{\pm}
\nonumber \\ & &
\pm i e ( \varphi^{A} 
        + \sfrac{1 - 2 s^{2}_{W}}{2 s_{W} c_{W}} \varphi^{Z} ) G^{\pm}
\enspace ,
\end{eqnarray}
where 
\begin{equation}
\label{eq:O-cs}  
O^{cs}_{h} := ( c_{\beta} O_{1h} + s_{\beta} O_{2h} )
\end{equation}
and the gauge parameters $\varphi^{a}$ are taken in physical directions.

For the full gauge transformations we have to transform the fermion 
and sfermion fields, too, since they transform under the same gauge 
groups as the higgs fields eq.(\ref{eq:gauge-h}). In the interaction 
basis, when we look at the unbroken Lagrangian, we know that the 
gauge transformations are \emph{flavour-blind}: they transform each 
generation independently in the same way. But once spontaneous 
symmetry breaking gives mass to gauge bosons, fermions, and sfermions
the Lagrangian has to be written in terms of these mass eigenstates
that no longer coincide with the interaction eigenstates. The mixing
of these states is described by the CKM matrix. In order to obtain the 
gauge-fixing and the ghost parts of the Lagrangian we write the 
gauge transformation in terms of the physical fields. But these 
\emph{new} gauge transformations are no longer \emph{flavour-blind}, as 
the $SU(2)$ gauge transformation mixes the up and down type quarks, which
are now superpositions of the different generations. And therefore
the full Lagrangian is no longer invariant under these \emph{new} gauge 
transformations. But since the $U(1)_{em}$ gauge group corresponding 
to the electric charge is unbroken, the full Lagrangian is still 
invariant under the $\varphi^{A}$ part of the gauge transformations.

\section{Non-linear gauge-fixing in FeynArts}
\label{sec:4}
\subsection{In the SM}
\label{sec:4.1}

From eqs.(\ref{eq:GZ},\ref{eq:GW},\ref{eq:Lgf}) it is easy to see, 
that we get non-standard interactions, that cannot be described by the
generic couplings of the SM or MSSM model files, which are defined in 
the file \texttt{Lorentz.gen} of FeynArts. Most of these non-standard 
interactions can be found in \texttt{Lorentzbgf.gen}, but there the 
ghost-ghost vertices are missing. They are needed for the mass- and 
field-renormalisation constants of the ghosts. So we modify the files in 
order to write a model file, that FeynArts can use to implement our 
non linear gauge fixing. 

\begin{table}
\caption{Field factors between our SM and FeynArts}
\label{tab:factors}       
\begin{tabular}{ll}
\hline\noalign{\smallskip}
fields & factor    \\
\noalign{\smallskip}\hline\noalign{\smallskip}
$Z_{\mu}$, $\bar{c}^{Z}$, $c^{Z}$ & $-1$  \\
$\chi^{\pm}$ & $\mp i$  \\
$\bar{c}^{A,Z,\pm}$ & $\xi_{A,Z,W}^{-1/2}$ \\
\noalign{\smallskip}\hline
\end{tabular}
\vspace*{1cm}  
\end{table}
For the SM our Lagrangian was obtained using different conventions than 
FeynArts. Therefore we had to redefine the fields according to 
table~\ref{tab:factors}. The modified Lorentz file and the model file 
can be downloaded from
\verb%http://terra.ar.fi.lt/~garfield/SM/ %.

\subsection{In the MSSM}
\label{sec:4.2}
For the MSSM we derive the Lagrangian using Superfields. We start 
from the unbroken supersymmetric Lagrangian, add the soft breaking 
terms and obtain the still gauge invariant full Lagrangian. After 
including spontaneous symmetry breaking and transforming the fields 
to the masseigenstates we add the gauge-fixing terms 
eq.(\ref{eq:Lgf}) and the ghost terms eq.(\ref{eq:Lgh}) that are 
calculated from the gauge fixing functions 
eqs.(\ref{eq:GA},\ref{eq:GZ-MSSM},\ref{eq:GW-MSSM}) with the gauge 
variation of the physical fields 
eqs.(\ref{eq:gauge:Amu}-\ref{eq:gauge:Gpm}). 

As an additional feature our model file includes the possible mixing 
of the two CP even with the CP odd neutral higgs bosons. With the new
FeynArts and FeynHiggs release of this summer, this feature is 
incorporated in the original distribution, too. When linking our 
model file for non-linear gauge-fixing, one has to set our mixing 
matrix \texttt{Ohiggs} to the FeynHiggs mixing matrices \texttt{Uhiggs}
or \texttt{Zhiggs}, which describe two different forms of mixings.

We use \texttt{Lorentzbgf.gen} as the basis for our modified Lorentz 
file. Since we choose our conventions in such a way as to be compatible
with FeynArts, we only had to add the ghost-ghost vertex to 
\texttt{Lorentzbgf.gen}. We create the model file for the MSSM with 
non-linear gauge-fixing, \texttt{MSSMnlgf.mod}, with the package 
ModelMaker, which is part of FeynArts. The Lorentz file and the model 
file can be downloaded from \\
\verb%http://terra.ar.fi.lt/~garfield/MSSM/%. 

The Mathematica programs for obtaining our MSSM-Lagrangian can be 
downloaded from the subdirectory \verb%Mathematica/%.
They have no documentation, though. 

\section{Outlook and Acknowledgements}
\label{sec:5} 

We want to do more checks of our model files. The analytic checks 
of the SM part could include also one-loop amplitudes, but for the MSSM
the counterterms to define a fully renormalised MSSM at one loop are 
still missing. It is also not clear, which renormalisation scheme
can be adopted, that allows general one loop calculations in the MSSM 
including general complex parameters. The normal on-shell scheme can 
not treat a decaying particle in an external line. The standard 
procedure \cite{denner} introduces the renormalisation condition 
$\widetilde{\mathrm{Re}}$, which cuts the absorptive parts in the 
selfenergy loops when determining the mass- and field-counterterms.
But when we calculate the decay-width of a decaying particle, we 
get a pole in our amplitude, when we do not include the width of the
particle in the propagator. This can be done systematically, using the 
complex mass scheme \cite{cms}. Since some people are still cautious 
about using a complex mass, we hope to apply the 
non-linear gauge-fixing to investigate the gauge independence of 
processes calculated with the complex mass scheme. 

If we can afford the time, we plan do provide a better documentation
to the calculation of the Lagrangian. 

In the long run we plan to calculate all counterterms in the MSSM in
a suitable framework, like the complex mass scheme, and include them 
in our model files. The idea behind this goal is to introduce particle
physics to the scientific community in Lithuania and to get students 
interested in this kind of work.

\subsection*{Thanks} 

We want to thank the SPA project \cite{SPA} for encouraging the open 
source idea and Thomas Hahn for the FeynArts support. T.G. thanks 
the HEPHY Vienna for the computer and the working place, where part of 
this work has been done and acknowledges the financial support of the 
VMSF, project Nr. T-07007.

%
%

\end{document}